\documentclass[useAMS,usenatbib,usegraphicx,useamssymb]{mn2e}
\usepackage{graphicx}
\usepackage{amssymb}

\graphicspath{{./}{images_peh/}}    
\title[]{Application of the Probability Event Horizon filter to constrain the local rate density of binary black hole inspirals with Advanced LIGO}
\author[]{E. Howell\thanks{E-mail: ejhowell@physics.uwa.edu.au}, D. Coward, R. Burman and D. Blair\\
 $^1$School of Physics, University of Western
Australia, Crawley WA 6009, Australia}

\date{Released 2002 Xxxxx XX}

\pagerange{\pageref{firstpage}--\pageref{lastpage}} \pubyear{2002}

\def\LaTeX{L\kern-.36em\raise.3ex\hbox{a}\kern-.15em
    T\kern-.1667em\lower.7ex\hbox{E}\kern-.125emX}

\begin{document}

\label{firstpage}

\maketitle

\begin{abstract}

The temporal evolution of the gravitational wave background signal resulting from stellar-mass binary black hole (BBH)
inspirals has a unique statistical signature. We describe the application of a new filter, based on the `probability
event horizon' (PEH) concept, that utilizes both the temporal and spatial source distribution to constrain the local
rate density, $r_{0}$, of BBH inspiral events in the nearby Universe. Assuming Advanced LIGO sensitivities and an upper
rate of Galactic BBH inspirals of $ 30\hspace{1mm}\mathrm{Myr}^{-1}$, we simulate GW data and apply a fitting procedure
to the PEH filtered data. To determine the accuracy of the PEH filter in constraining $r_{0}$, a comparison is made
with a fit to the brightness distribution of events. We apply both methods to a data stream containing a background of
Gaussian distributed false alarms. We find that the brightness distribution yields lower standard errors, but is biased
by the false alarms. In comparison the PEH method is less prone to errors resulting from false alarms but has a lower
resolution as fewer events contribute to the data. Used in combination, the PEH and brightness distribution methods
provide an improved estimate of the rate density.

\end{abstract}


\section{Introduction}
The LIGO gravitational wave (GW) detectors are currently taking data at design sensitivity, and embarking on
long science runs. Promising GW sources potentially detectable by LIGO are coalescing binary systems
containing neutron stars (NSs) and/or black holes (BHs)-- see \cite{FH98}, \cite{MGS04}, \cite{Thorn95}.
Because of the enormous GW luminosity $\sim 10^{-3} c^{5}/G \sim 10^{23}L_{\odot}$, binary black hole (BBH)
inspirals  are among the most promising candidates for a first detection of GWs
\citep{Baker02,FH98,Saty04,burger05}.

For LIGO-type detectors, even highly energetic BBH inspirals are predicted to be detected at a rate of only
$(10^{-3}-0.6)\hspace{0.5mm}\mathrm{yr}^{-1}$ \citep{CT02}. However, the next generation of interferometric detectors,
planned to go online in the next decade, should be sensitive to an abundance of sources, with event rates 1000 times
greater than current detectors. These `Advanced' interferometers will provide a new window to the cosmos not accessible
by conventional
astronomy.\\
\indent The introduction of advanced detectors will allow us to detect BBH inspiral events out to a distance of $z
\approx 0.4$. Events within this volume will contribute to the low probability `popcorn' component of the astrophysical
GW background for BBHs. It has been shown that the temporal and spatial distribution of transient sources which form this
part of the GW
background can be described by the Probability Event Horizon (PEH) concept of \cite{CB05}.\\
\indent The PEH describes the temporal evolution of the brightness of a class of transient events. The probability of a
nearby event accumulates with observation time, so that the peak event amplitude has a statistical distribution
dictated by the rate density and spacetime geometry. This feature provides a tool to model the detectability of a
distribution of transient GW sources. \cite{Coward_NSM05} used the PEH to model the detectability of NS inspirals as a
function of observation time assuming LIGO and Advanced LIGO
sensitivities and reasonable estimates of the local rate density of events in the nearby Universe, $r_{0}$.\\
\indent In this study we consider the use of the PEH method to determine the local rate density of BBH coalescence
events. We use a cosmological model to create synthetic data corresponding to four months data at advanced LIGO
sensitivity. We then attempt to recover the
assumed rate density.\\
\indent First we use the conventional approach of studying the brightness distribution of all events. This method does
not make use of the time evolution. It simply utilizes the number-amplitude distribution and fits the observations to
the rate density. We compare this with the PEH method, which utilizes the temporal distribution of events. For
simplicity, we use a standard candle approximation to model our source population. We note that a network of detectors
would allow us to exploit a special property of compact binary inspiral events, for which the \emph{chirp} signal
provides a measure
of the luminosity distance to the source, enabling such sources to be treated as standard candles \citep{schutz86,Saty04,CF93, Finn93}.\\

\indent To test both methods, we simulate a candidate population of BBH inspiral events by approximating the output
data stream resulting from matched filtering. A simplified detection model (described in Section 5.1) is used which
assumes Gaussian detector noise. We use the candidate population to show that the temporal evolution of events has a
unique statistical signature and exploit this signature to constrain $r_{0}$. To determine the effect of detector
efficiency we apply both methods to data corresponding to high and low false alarm rates. We also consider the effect
of different
star formation rate (SFR) models and determine bias introduced by different SFR evolution functions.\\
\indent The paper is organized as follows: In Section 2 we review event rate predictions  and then discuss the
evolution of the event rate of BBH inspirals in Section 3. In Section 4, we explain the PEH concept and show how it can
be used to probe the source rate density and to differentiate between different astrophysical populations of sources.
We describe the simulation of candidate events in Section 5 and use these data in Section 6 to determine an estimate of
$r_{0}$ using the brightness distribution of sources. In Section 7 we present a method for extracting PEH data and use
least-squares fitting to constrain $r_{0}$. We present our results in Section 8 and in Section 9 summarize the key
findings and discuss how this work can be extended.

\section{EVENT RATE ESTIMATIONS}

The local rate density of a particular astrophysical GW source is fundamental to estimating the number of potentially
observable events. Usually defined within a volume spanning the Virgo cluster of galaxies, the local rate density,
$r_{0}$, is determined using estimated source rates within a larger fixed volume of space.

In the case of NS-NS inspirals, current rate estimates rely on a small sample of sources. The discovery of the double
pulsar PSR J0737\hspace{0.5mm}--\hspace{0.5mm}3039, with estimated coalescence time $\sim 87$ Myr, increased the
estimated inspiral rate for double NSs in our Galaxy by about an order-of-magnitude to $20  - 300
\hspace{1mm}\mathrm{Myr}^{-1}$ \citep{ burgay03, kalogera04} with respect to earlier estimates \citep{kalogera01,
phinney91}.

The rates of BBH systems are even more uncertain. Because these systems have not been observed directly, their
evolutionary parameters can be obtained only though population synthesis, which predicts Galactic event rates $\le
10^{-1}-80\hspace{0.5mm}\mathrm{Myr}^{-1}$. However, GW emissions from BBH will be detectable out to much greater
distances than other systems of coalescing compact objects \citep{Saty04}. The rates of BH-NS inspirals are of a
similar range to those of BH - BH systems, but have lower expected detection rates as a result of their less energetic
emissions.

In this study, we use the Galactic BBH coalescence rate ${\cal R}_{\rm gal}^{BBH}\sim 30
\hspace{1.0mm}\mathrm{Myr}^{-1}$, obtained from the standard model in the population synthesis calculations of
\cite{Belczynski02}. We note that this rate is an upper limit. We convert this to a rate per unit volume, $r_{0}$,
using the conversion factor $10^{-2}$ from \cite{Ando04} for the number density of galaxies in units of
$\mathrm{Mpc}^{-3}$, yielding the reference value of the local rate density, $\tilde{r}_{0} =
0.3\hspace{1.0mm}\mathrm{Myr}^{-1}\hspace{0.5mm}\mathrm{Mpc}^{-3}$, which we will employ in this study.

Rate estimates will benefit greatly from the introduction of advanced GW detectors, which will allow unobscured source
counts to be conducted to almost cosmological volumes. A network of detectors will improve the sky coverage and source
localization. A network may employ coherent analysis, in which synchronized detector outputs are merged \citep{Finn01}
before a search for a common pattern, or alternatively, a coincidence analysis, in which individual events from
different detectors are correlated in time \citep{Arnaud02}. However, as a result of the non-uniform antenna patterns,
even a network of three detectors of similar sensitivity will have difficulty obtaining maximum efficiency
\citep{Arnaud03a}. In addition, the efficiency of a detector network for a particular source type will depend on the
false alarm rate and the signal-to-noise threshold for detection -- therefore high number counts may be balanced by an
increased rate of false alarms.  We will investigate this in Section 6.

\section{THE BBH COALESCENCE RATE EVOLUTION}

\subsection{The event rate equation}

For standard Friedman cosmology a differential event rate in the redshift shell $z$ to $z+{\mathrm d}z$ is
given by:

\begin{equation}\label{drdz}
\mathrm{d}R = \frac{\mathrm{d}V}{\mathrm{d}z}\frac{r_0 e(z)}{1+z} \mathrm{d}z \,,
\end{equation}

\noindent where $\mathrm{d}V$ is the cosmology-dependent co-moving volume element and  $R(z)$ is the all-sky ($4\pi$
solid angle) event rate, as observed in our local frame, for sources out to redshift $z$. Source rate density evolution
is accounted for by the dimensionless evolution factor $e(z)$, normalized to unity in our local intergalactic
neighbourhood, and $r_0$ is the $z=0$ source rate density. The $(1 + z)$ factor accounts for the time dilation of the
observed rate by cosmic expansion, converting a source-count equation to an event rate equation.

The cosmological volume element is obtained by calculating the luminosity distance from (cf. \cite{Peebles},
p.\hspace{0.5mm}332)

\begin{equation}\label{lumd}
d_\mathrm{L}(z) = (1 + z)\frac{c}{H_{0}}\int_{0}^{z} \frac{\mathrm{d}z\hspace{0.5mm}'}{h(z\hspace{0.5mm}')}\,,
\end{equation}

\noindent and using (Porciani and Madau 2001, eq.\hspace{1.0mm}3)

\begin{equation}\label{dvdz}
\frac{\mathrm{d}V}{\mathrm{d}z}= \frac{4\pi c}{H_{0}}\frac{d_\mathrm{\hspace{0.25mm}L}^{\hspace{1.5mm}2}(z)}{(1 +
z)^{\hspace{0.25mm}2}\hspace{0.5mm}h(z)}\hspace{2.0mm}.
\end{equation}

\noindent The normalized Hubble parameter, $h(z)$, is given by

\begin{equation}\label{hz}
h(z)\equiv H(z)/H_0 = \big[\Omega_{\mathrm m} (1+z)^3+ \Omega_{\mathrm \Lambda} \big]^{1/2}\,
\end{equation}

\noindent for a `flat-$\Lambda$' cosmology $(\Omega_{\mathrm m} + \Omega_{\mathrm \Lambda}=1$). We use $\Omega_{\mathrm
m}=0.3$ and $\Omega_{\mathrm \Lambda}=0.7$ for the $z\hspace{-0.05cm}=\hspace{-0.05cm}0$ density parameters, and take
\mbox{$H_{0}=70$ km s$^{-1}$ Mpc$^{-1}$} for the Hubble parameter at the present epoch.

\subsection{The source rate evolution of BBH}

For the source rate evolution factor, $e\hspace{0.3mm}(z)$, we employ three star formation rate (SFR) models and a
non-evolving SFR density for comparison. To simulate a candidate population of BBH inspiral events we employ the
observation-based SFR model SF2 of \cite{PM01}. Based on observed rest-frame ultraviolet and H$\alpha$ luminosity
densities, this model includes an allowance for uncertainties in the amount of dust extinction at high $z$. In order to
constrain $r_{0}$ by least-squares fitting to the simulated data, we will use three additional models: a non-evolving
SFR density model obtained by setting $e(z) = 1$, the model SF1 of \cite{PM01} and the model SH, based on an analytical
fit to hydrodynamic simulations conducted by \cite{SH03} in a flat-$\Lambda$ cold dark matter cosmology; SF1 includes
an upward correction for dust extinction at high $z$. We re-scale SF1 and SF2, originally modelled in an
Einstein-de-Sitter cosmology, to a flat-$\Lambda$ cosmology using the procedure outlined in the appendix of Porciani \&
Madau.

\begin{figure}
\includegraphics[width=84mm]{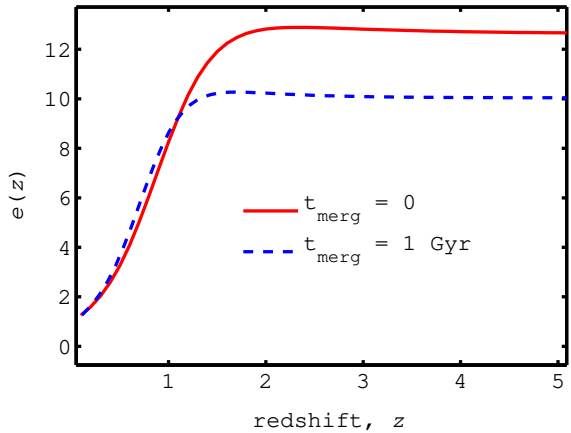}
\caption{The solid line shows the dimensionless SFR density evolution factor $e(z)$ for the SFR model SF2 of
\citet{PM01} in the flat-$\Lambda$ (0.3,\hspace{1.0mm}0.7) cosmology. To allow for the average coalescence
time of BBH systems, the dashed line shows the effect of a time delay of 1 Gyr on $e(z)$.} \label{fig_ez}
\end{figure}

\begin{figure}
\includegraphics[width=84mm]
{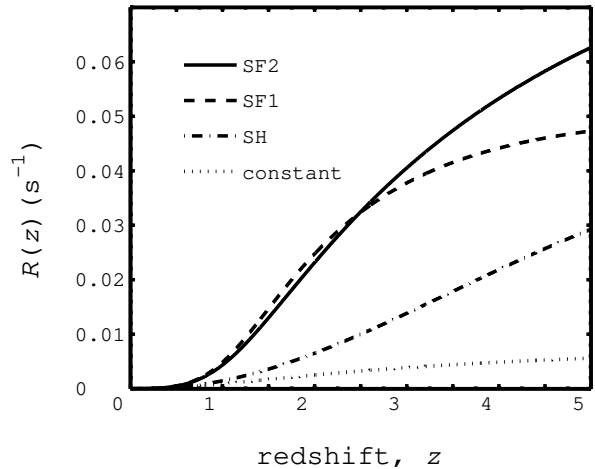}
 \caption{The all-sky BBH coalescence rate as a function of $z$ using a Galactic rate
 ${\cal{R}}_{\mathrm{Gal}}^{BBH}\sim 30 \hspace{1.0mm}\mathrm{Myr}^{-1}$ and merger time of 1 Gyr
 (Belczynski et al. 2002). We employ the observation-based star formation rate models SF1 and SF2
 of \citet{PM01}, a constant (non-evolving) model and
 a simulation-based model, SH, of \citet{SH03}.}
  \label{fig_rz}
\end{figure}

\begin{figure}
\includegraphics[width=84mm]{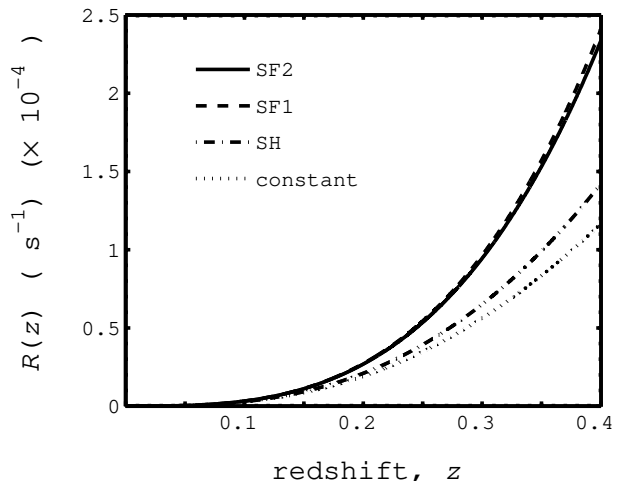}
 \caption{The all-sky rate of BBH coalescence as a function of $z$ as previously shown in Figure \ref{fig_rz}, but
 within a distance of $z=0.4$, corresponding to the Advanced LIGO detectability horizon for these sources.}
  \label{fig_rz_low}
\end{figure}

\cite{Coward_NSM05} assumed that the formation of double NS systems closely tracks the evolving star formation rate.
They based their assumptions on short merger times and showed that a time delay of up to 5 Gyr had minimal effect on
the differential rate of events at low redshift. However, in comparison with NS-NS systems, for which the distribution
of merger times has a large range, peaking at around 0.3\hspace{0.5mm}Myr \citep{Belczynski02}, the merger time
distributions of BBH systems are predominantly skewed towards longer merger times -- from about 100 Myr to the Hubble
time \citep{BBR04}.

Belczynski et al. used population synthesis methods to calculate the properties and coalescence rates of
compact binaries. They used a range of different scenarios defined by the initial physical parameters of the
binary system such as component masses, orbital separations and eccentricities. They also included
properties which affect the evolutionary channels of the system, such as mass transfer, mass losses due to
stellar winds and kick velocities. Their calculations implied that, compared to NS-NS systems, the wider
orbits and stronger dependence of merger time on initial separation for BH-NS and BH-BH systems resulted in
longer merger times, mostly in the range 0.1 to several Gyr (Belczynski et al. section 3.4.5).

A standard model was defined by \cite{Belczynski02} using a range of assumptions, including: a
non-conservative mass transfer with half the mass lost by the donor lost by the system; a kick velocity
distribution that accounts for the fact that many pulsars have velocities above 500 km $\mathrm{s}^{-1}$;
constant Galactic star formation for the last 10 Gyr. For this model they found a distribution in BBH merger
times that peaked at~\hspace{0.5mm}1 Gyr. We therefore take this value as an average merger time and shift
$e\hspace{0.3mm}(z)$ to reflect this delay time. Using the method described in section 2 of
\cite{Coward_NSM05}, we can convert $e(z)$ to a function of cosmic time, $t_{\mathrm{cos}}$, using the
relation

\begin{equation}\label{eq_cos_time}
t_{\mathrm{cos}}(z) =
\int_{0}^{z}[\hspace{0.5mm}(1+z\hspace{0.5mm}')h(z\hspace{0.5mm}')\hspace{0.5mm}]^{-1}\hspace{0.5mm}dz\hspace{0.5mm}'\hspace{0.5mm}.
\end{equation}

\noindent We apply a 1 Gyr time shift and then convert back to a function of $z$. Figure \ref{fig_ez} shows the factor
$e(z)$ for SF2 with and without the time delay. The result shows that within the range of advanced LIGO, $z \sim 0.4$,
$e(z)$ is not significantly altered. Therefore, although we include this time delay in our calculations, within the
ranges of advanced LIGO detectors it will not have a significant influence on the results.

Figure \ref{fig_rz} shows the all-sky BBH coalescence rate, $R\hspace{0.5mm}(z)$, calculated by integrating the
differential rate from the present epoch to redshift $z$. Using SF2, the rate continues increasing to distances well
beyond the cosmological volume elements considered in this study. We therefore assume a universal rate of events,
$R^{\mathrm{U}}$ (usually defined as the asymptotic value of the all-sky rate as $z$ increases) of $\sim
0.06\hspace{0.1cm} \mathrm{s}^{-1}$ and a corresponding mean temporal interval $\tau = 1/R^{\mathrm{U}} =
17$\hspace{0.5mm}s.

Figure \ref{fig_rz_low} shows the all-sky BBH coalescence rate for sources within Advanced LIGO sensitivities. The
potential detection horizon extends to $z = 0.4$ and the corresponding mean rate of events is $\sim 2.5\times
10^{-4}\hspace{1mm}\mathrm{s}^{-1}$ using SF2. If we assume all sources are composed of two $10 M_{\odot}$ black holes,
this rate corresponds to around 874 events $\mathrm{yr}^{-1}$ with SNR $\ge 8$.

\section{THE PROBABILITY EVENT HORIZON FOR BBH COALESCENCE EVENTS}

\subsection{The Probability Event Horizon}

The rate, as observed in our frame, of transient astrophysical events occurring throughout the Universe, is determined
by their spatial distribution and evolutionary history. The distribution of event observation times follows a Poisson
distribution and the temporal separation between events follows an exponential distribution defined by the mean event
rate. For cosmological events, the rate depends on the cosmology dependent volume and radial distance through redshift,
$z$. We assume that an observer measures both a temporal location and a `brightness' for each event, where the
brightness is determined by the luminosity distance to the event.

It follows from these assumptions that the probability for at least one event to occur in the volume bounded by $z$,
during observation time $T$ at a mean rate $R(z)$ at constant probability $\epsilon$ is given by the exponential
distribution:

\begin{equation}\label{prob2}
p(n\ge1;R(z),T) = 1 - e^{-R(z) T}= \epsilon\,, \label{eq_peh}
\end{equation}

\noindent  $1 - e^{-RT}$ being the probability of at least one event occurring (see Coward \& Burman 2005).

For equation (\ref{prob2}) to remain satisfied as observation time increases, the mean number of events in the sphere
bounded by $z$, $N_{\epsilon}= R(z)T=\vert \mathrm{ln}(1-\epsilon)\vert$, must remain constant. The PEH is defined by
the redshift bound, $z_{\epsilon}^{\mathrm{PEH}}(T)$, required to satisfy this condition.

We note that the for events occurring within a volume bounded by about a few Gpc, the PEH is well approximated using
Euclidean geometry and takes on a simple analytical form with $z$ now replaced by the radial distance in flat space:

\begin{equation}\label{horizon}
r_{\epsilon}^{\mathrm{PEH}}(T)= (3N_{\epsilon}/4\pi r_0)^{1/3} T^{-1/3}\;, \label{eq_peh_euc}
\end{equation}

\noindent where $r_0$ is a rate per unit volume. By setting $\epsilon = 0.95$, one can generate a threshold
corresponding to a 95\% probability of observing at least one event within $z_{\epsilon}^{\mathrm{PEH}}(T)$, or
alternatively, $r_{\epsilon}^{\mathrm{PEH}}(T)$ for a Euclidean PEH model. We define a `null PEH', representing the
95\% probability that no events will be observed within this threshold, by setting $\epsilon = 0.05$. When combined, we
refer to these two PEHs as the 90\% PEH band -- that is 90\% of events are expected to occur in the region enclosed by
the two PEHs. By scaling some fiducial GW amplitude by $d_\mathrm{L}(z)$, we  express the 90\% PEH band as 90\%
confidence bounds of peak GW amplitude against observation time.

\begin{figure}
\includegraphics[width=84mm]{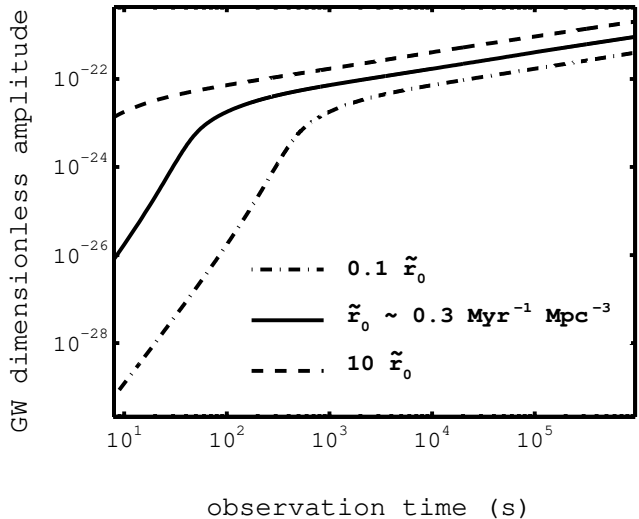}
 \caption{The GW amplitude 95\% PEH curves for BBH using three values for the local rate density,
 $r_{0}$. The PEH curve corresponding to the reference value of the local rate density, $\tilde{r}_{0}$,
 used in this study is shown by the solid line.
 An increase in $r_{0}$ by an order of magnitude shifts the PEH curve towards earlier observation
 times (dashed line). This implies an increased probability for detecting a local high-amplitude event. The dot-dashed line
 shows that a decrease in  $r_{0}$ of the same magnitude has the reverse effect.}
  \label{fig_peh_lrd}
\end{figure}

Figure \ref{fig_peh_lrd} shows the Euclidean 95\% PEH curves for BBH, using three different values of $r_0$, differing
by an order of magnitude. We note that for a particular transient GW source population, the PEH is intrinsically
dependent on $r_0$. The plot shows that for a particular source type, the PEH model can be used to estimate the value
of $r_0$.

\begin{figure}
\includegraphics[width=84mm]{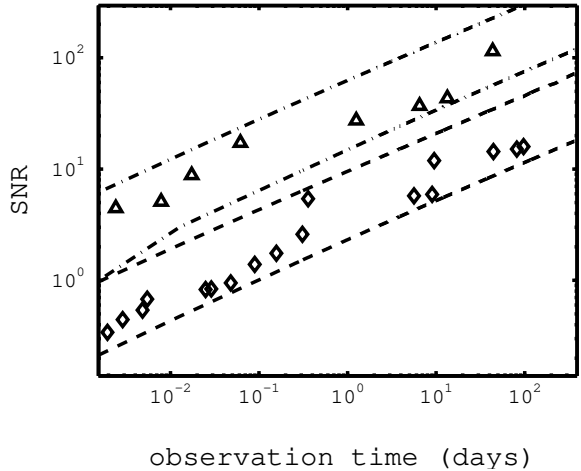}
 \caption{To illustrate that our simulated data is consistent with the PEH model, we show the 90\% PEH
 thresholds (see Section 4.1) and simulated PEH data for two different GW binary inspiral source types with
 varying event rates recorded within a redshift $z$. The PEH thresholds show the optimal SNR
 for an Advanced LIGO detector as a function of observation time. We assume a
 standard-candle approximation based on the fiducial distances for an optimal SNR of 8 for each source.  The fiducial distances and universal event rates for
  the GW populations are: NS-NS inspirals
  at 200 Mpc and $R^{U}\sim 0.2  \hspace{1mm}\mathrm{s}^{-1}$(represented by diamonds) and BBH at 2 Gpc and
  $R^{U}\sim 0.05  \hspace{1mm}\mathrm{s}^{-1}$(indicated by triangles).}
  \label{fig_peh_sources}
\end{figure}

\subsection{The PEH filter applied to astrophysical populations}

\cite{Coward_NSM05} outlined the PEH filter -- a procedure used to extract a cosmological signature from a distribution
of events in redshift and time. The concept is very simple: the longer one observes the greater the probability of a
nearby event -- the PEH filter quantifies this dependence as a means of probing the cosmological rate density. The PEH
filter searches for the time dependence of event amplitudes which is imposed by their cosmological distribution. It is
a non-linear filter applied to a body of data by recording successively closer events. We will apply the technique to a
distribution of amplitudes, $A$, as a function of observation time, $t$, recording the successive $(t_{i},A_{i})$ that
satisfy the condition $A_{i+1}> A_{i}$. The resulting events will be referred to as the `PEH population'.

By fitting to PEH data we can probe the local rate density of an astrophysical population. We note that just as
brightness distributions can be used to separate and identify different source populations, so also can different source
populations be identified in PEH plots.


Figure \ref{fig_peh_sources} demonstrates this property by comparing synthetic PEH populations of BBH and NS-NS
inspirals. For the two source types we use a standard candle approximation based on an optimal SNR of 8 assuming an
Advanced LIGO detector with a fiducial distance of 200 Mpc for NS-NS inspirals and 2 Gpc for BBH inspirals.

We show the 90\% PEH bands for the two populations as a SNR. It is evident that if the PEH signature of an
astrophysical GW background could be extracted from detector noise, the presence of different astrophysical backgrounds
could be identified. Clearly this method will only work if the luminosities of the populations differ widely. In
reality, different luminosities will also be associated with different waveforms, and there are likely to be more
evident differences than the simple luminosity effects.

\section{SIMULATING A CANDIDATE POPULATION OF BBH INSPIRALS}

\subsection{The GW source model}

\indent We will consider only the well understood inspiral stage of a BBH coalescence to provide quantitive data for
our source model. To model the GW background from BBH inspirals it is sufficient to assume that the raw interferometer
data is preprocessed by passing it through an optimal filter. This transforms events into approximate short duration
Gaussian pulse signals \citep{Abbott04, SO04} embedded in Gaussian detector noise. By injecting a population of
simulated events into GW detector noise of Advanced LIGO sensitivity, we approximate the processed output data stream
of a GW detector and then apply sub-optimal burst filtering to extract candidate events or `triggers'. A more realistic
detection pipeline will employ a range of templates representing different BBH mass configurations. To simplify our
detection model, we assume a single BBH mass configuration for our sources, representing the output of a fixed
template.

\begin{figure}
\includegraphics[bbllx = 11pt,bblly =79pt, bburx = 600 pt, bbury = 739pt,angle = 0,scale = 0.45, origin=lr]{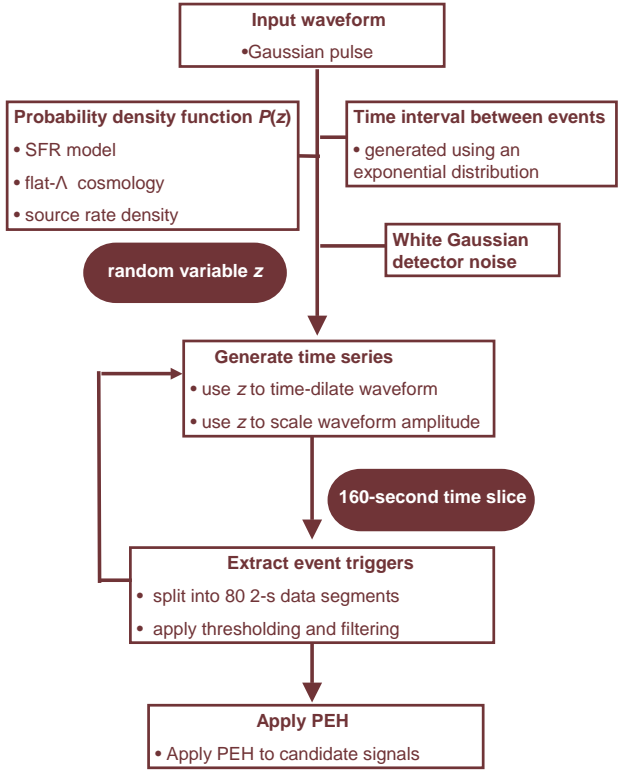}
 \caption{Flow chart outlining the simulation pipeline to generate a candidate population of
BBH inspiral events in interferometer detector noise. Individual inspiral events are scaled in amplitude and
time-dilated according to the random variable $z$, obtained from the probability distribution shown in
Figure \ref{fig_pdf}. The events are injected into simulated detector noise, with the temporal separations
between successive events following an exponential distribution. Candidate signals are extracted using
amplitude thresholding and robust sub-optimal filtering.} \label{fig_pipeline}
\end{figure}

\begin{figure}
\includegraphics[width=84mm]{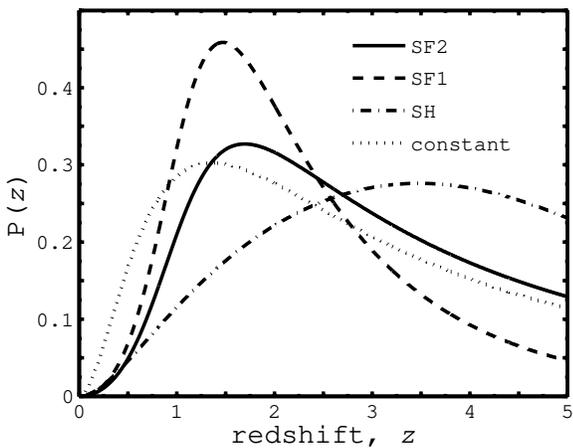}
 \caption{The probability distribution function for BBH inspirals throughout the Universe modelled using the same parameters as for
 the cumulative rate (see Fig. \ref{fig_rz}). For SF1, SF2 and the constant (non-evolving) model, the most probable events occur in $z \approx 1 - 2$.
 For SH, the most probable events occur significantly earlier, in $z \approx 3 - 4$, but with a flatter distribution.}
  \label{fig_pdf}
\end{figure}

To represent the response to BBH events by optimal matched filtering, we adopt as a model waveform the simplified but
robust form used by \cite{Arnaud03} and \cite{Abbott05} to model GW burst sources. This is a linearly polarized
5\hspace{ 0.5mm}-ms duration Gaussian pulse, which approximates to the form of the event triggers in processed data
from the LIGO S1 search for inspirals \citep{Abbott04,SO04}. We note that we have chosen to ignore any additional
secondary peaks which occur at high SNRs.

The two polarizations of this signal will be given by:

\begin{equation}
h_{+}(t)=A\hspace{0.5mm}\mathrm{exp}\left[{-\frac{\left(\textstyle{t}-
\textstyle{t}_{0}\right)^{2}}{\textstyle{2}\textstyle{\Delta}^{2}}}\right],\hspace{2.0mm} h_{\times}(t) =
0\hspace{0.5mm} \label{eq-pdf}\hspace{0.25cm}\label{eqn_pulse}
\end{equation}

\noindent with amplitude $A$ and half-width $\Delta$; the time value of the signal maximum, $t_{0}$, is set to
10$\hspace{0.5mm}$ms. The output response, $h(t)$, will be a linear combination of the two polarizations

\begin{equation}
h(t)= F_{+}h_{+} + F_{\times}h_{\times}\hspace{0.5mm}, \label{eq_antenna}
\end{equation}

\noindent where $F_{+}$ and $F_{\times}$ are the antenna pattern functions, which are functions of sky
direction, represented by the spherical polar angles $\theta$ and $\phi$, and polarization angle, $\psi$, of the
GW signals relative to the detector \citep{Jaranowski98}.

The filter response amplitude, $A$, will be dependent on the masses of the BBH system. For simplicity, rather than
using a distribution of BBH masses, we use a standard source model. Using the \emph{BENCH} \footnote{The program BENCH
can be obtained from the URL http://ilog.ligo-wa.caltech.edu:7285/advligo/Bench} code to model the detector noise
spectrum, we approximate the GW amplitude for an optimum SNR of 80 at 200 Mpc.

By assuming a standard BBH system we ignore the amplitude distribution from the BH mass spectrum. However, this
assumption is not a limitation to the PEH method because the signature of an inspiralling binary system contains a
measure of its luminosity distance \citep{schutz86, Saty04, Finn93, CF93}. Therefore, for a network of GW detectors,
the analysis described in this paper could be repeated using luminosity distances rather than amplitudes.

\subsection{Simulation of GW interferometer data}

The simulation pipeline used to generate a cosmological GW population of BBH inspiral triggers in GW detector noise is
shown in Figure \ref{fig_pipeline}. We use the \emph{BENCH} code to calculate the noise sensitivity curve corresponding
to our source model.

Following \cite{Arnaud03}, we define the standard deviation of the detector noise as:

\begin{equation}
\sigma = h_{\mathrm{rms}}\sqrt{f_{\mathrm{o}}/2 f_{\mathrm{c}}}\hspace{0.5mm}, \label{eq_sigma}
\end{equation}

\noindent with $h_{\mathrm{rms}}$ the root-mean-square value of the advanced LIGO noise curve at rest frame frequency
$f_{\mathrm{c}}$, and $f_{\mathrm{o}}$ the sampling frequency. We assume the noise is white Gaussian with zero mean.

The amplitude and duration of each potential BBH inspiral event is defined by the random variable $z$, generated from a
probability density function $P(z)$ for these events. We obtain $P(z)$ by normalizing the differential event rate,
$\mathrm{d}R/\mathrm{d}z$, by $R^{U}$, the Universal rate of BBH inspiral events, integrated throughout the cosmos, as
seen in our frame (cf. Coward \& Burman 2005, section 3):

\begin{equation}
P\hspace{0.5mm}(z)\hspace{0.5mm}{\mathrm d}z = {\mathrm d}R/R^{\mathrm{U}} \;.\label{Pz}
\end{equation}

Equation \ref{Pz} defines $P(z)\hspace{0.5mm}\mathrm{d}z$ as the probability that an observed event occurred in the
redshift shell $z$ to $z + \mathrm{d}z$. In Figure \ref{fig_pdf} we present curves for the several star formation rate
models. We see that the most probable events will occur at $z\approx 1-2$ for SF1 and SF2. The corresponding cumulative
distribution function $C(z)$, giving the probability of an event occurring in the redshift range $0$ to $z$, is the
normalized cumulative rate\hspace{0.5mm}:

\begin{equation}
C(z) = R(z)/R^{\mathrm{U}} \;. \label{cz}
\end{equation}

We use $C(z)$ to simulate events and their associated redshifts and hence the GW amplitude of each injected
candidate event, $h(t)$, is computed from equation (\ref{eq_antenna}). Random values of the antenna response
variables, $\theta$, $\phi$ and $\psi$, are simulated and $h(t)$ is scaled inversely by the luminosity
distance $d_{\mathrm{L}}(z)$. The signal duration is time-dilated by the factor $(1 + z)$.

The temporal distribution of events in our frame is stochastic and the separation of events is described by Poisson
statistics. The time interval between successive events, $\tau$, will therefore follow an exponential distribution.
Successive waveforms, with amplitude and distance defined by the random variable $z$ are generated and injected into a
data pipeline at time intervals $\tau$.

\begin{figure}
\includegraphics[width=84mm]
{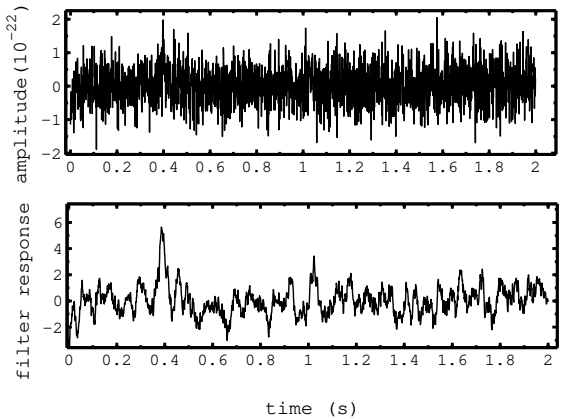}
 \caption{The top panel shows an optimally orientated burst signal \mbox{(a 5-\hspace{0.5mm}ms} Gaussian pulse)
 at $t \approx 0.4$\hspace{0.5mm}s with an optimal SNR of 8 when embedded in Gaussian noise corresponding to 200 Hz of
the Advanced LIGO noise spectrum. The bottom panel shows the response of the mean filter in terms of a SNR. The mean
filter response to this signal is about 70\% of that of an optimal filter.}
  \label{fig_filter}
\end{figure}

\begin{figure}
\includegraphics[width=84mm]
{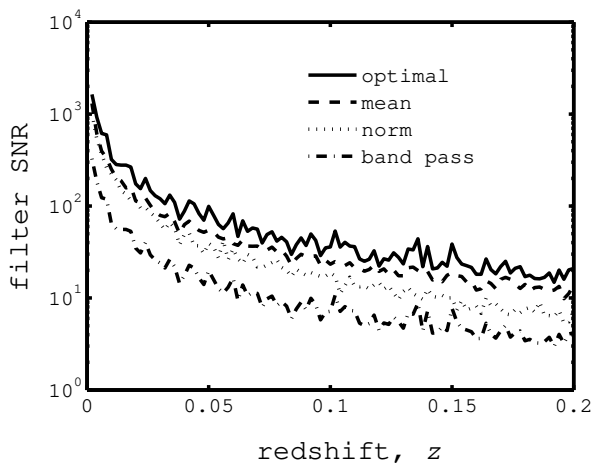}
 \caption{The effective SNR performance of three GW burst filters, as functions of $z$, in identifying the optimally
 orientated Gaussian pulse signal shown in Fig. \ref{fig_filter}. Compared with the performance of an optimal
 filter in Gaussian detector noise with Advanced LIGO sensitivity, the mean and norm filters provide useful performance.
Noise samples are generated for each $z$ -- SNR fluctuations are the
 result of RMS variations between noise samples. }
  \label{fig_filterSNR}
\end{figure}

\subsection{Candidate signal extraction}
We restrict ourselves to the well understood coalescence phase of BBH sources and as discussed earlier matched
filtering provides processed data whereby candidate events can be approximated as Gaussian bursts embedded in a
background of noise \citep{Abbott04, SO04}. Candidate events will usually be selected on the basis of coincidence
analysis and a $\chi^{2}$ waveform consistency test performed between template and filtered output \citep{Abbott04}.

In our simulation pipeline we inject signals directly into Gaussian detector noise, thereby contaminating potential
triggers; this makes a simple thresholding procedure insufficient to extract good candidates. A combination of matched
filtering and `pulse' detection techniques such as `burst filters' has been suggested by \cite{Pradier01} as a means of
increasing the final SNR for GW inspiral events. We therefore employ burst filtering to extract a population of event
candidates.

For simplicity, we employ the mean filter, a highly robust, linear filter developed by \cite{Arnaud03}, which operates
in the time domain by calculating the mean of the data, $x_{i}$, in a sliding window of sample width $N$:

\begin{equation}
\hat{y}_{\mathrm{j}} = \frac{1}{N}\sum_{\mathrm{i}=\mathrm{j}}^{\mathrm{j} + N - 1}x_{\mathrm{i}}\hspace{0.3mm}.
\end{equation}

Figure \ref{fig_filter} shows the response of this filter to an optimally orientated Gaussian pulse (see equation
\ref{eqn_pulse}) at a distance of $z = 0.4$, just within the expected detection limit for Advanced LIGO \citep{Saty04,
CT02}. The response is displayed as a maximum SNR -- the maximum value of the ratio of mean filter output when a signal
is present to the standard deviation in the absence of a signal \citep{Arnaud03,FH98}. The maximum response of about
5.5 is around 70\% of that using optimal filtering -- for which a SNR of about 8 was obtained. This mean filter
response is typical of values obtained during testing.

Figure \ref{fig_filterSNR} shows the comparative filter performances obtained for our pulse model at
different values of $z$ operating in white Gaussian noise, comparable in amplitude to that of the 200 Hz
region of the Advanced LIGO noise curve. We compare the responses of the mean filter, norm filter and a
simple band-pass filter, with that of a Wiener filter, with optimal SNR given by

\begin{equation}
\rho_{0}^{2}=4\int_{0}^{\infty}\frac{|\tilde{h}(f)|^{2}}   {   \mathrm{S}_{    \mathrm{h}   } (f)
}\hspace{1.0mm}\mathrm{d}f\hspace{0.3mm},
\end{equation}

\noindent with $\mathrm{S}_{\mathrm{h}}$ denoting the one-sided noise power spectral density. We see that
the mean filter is the sub-optimal filter with the best average response. The fluctuations in the response
are a result of generating different noise samples for each $z$. These results are in agreement with tests
on time domain filters carried out by \cite{Arnaud03} and \cite{Bizouard03}. The performance level, coupled
with the robustness of this filter, make the mean filter an ideal choice for our candidate searches.

The simulation pipeline for the generation of candidate events consists of the following steps: \\

\noindent\textbf{1.} A 160\hspace{0.2mm}-s data buffer, representing the output $\rho(t)$ of a single optimal filter,
is continuously populated with Gaussian detector noise at a sampling frequency of
1024 Hz --- in an actual application, this represents a down-sampling of interferometer data from $2^{14}$ Hz.\\

\noindent\textbf{2.} Potential candidate events corresponding to different cosmological distances are injected into the
data buffer at exponentially distributed intervals $\tau$ as described in section 5.2. These intervals are recorded.\\

\noindent \textbf{3.} The 160\hspace{0.2mm}-s data segments are split into 80$\times $2\hspace{0.5mm}s slices.\\

\noindent \textbf{4.} If a slice contains no fluctuations above a threshold of $\sigma = 3.2$, it is rejected.
Otherwise, a mean filter is applied to the slice -- an event with maximum SNR $>$
4.2 is recorded as a candidate BBH inspiral event, and is added to a candidate event population, $E$.\\

\noindent \textbf{5.} The PEH algorithm is applied to the candidate events to extract the consecutive running
maximum amplitudes -- these events we describe as the candidate PEH population, $C$.\\

Both event triggers and injected signals are recorded for later analysis.

\subsection{The candidate event population}

\begin{figure}
\includegraphics[width=84mm]
{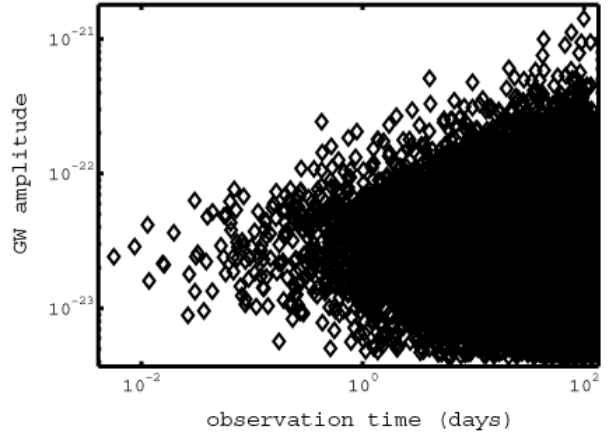}
 \caption{The simulated BBH inspiral population assuming a local rate density of
 $0.3\hspace{0.5mm}\mathrm{Myr}^{-1}\hspace{0.5mm}\mathrm{Mpc}^{-3}$  for 4 months of data, corresponding
 to 527,680 events. Note how the GW amplitudes incorporate both higher and lower magnitude extremes as observation time
 increases -- corresponding to smaller and \mbox{larger $z$.}}
  \label{fig_injected}
\end{figure}

Figure \ref{fig_injected} shows the injected events representing the GW background of BBH within $z \sim 5$ for four
months of observation time, corresponding to around 500,000 events --- about 2200 of these events were within $z =
0.4$, the detectabilty horizon of Advanced LIGO, a value which is within the upper limit of
8000\hspace{0.5mm}$\mathrm{yr}^{-1}$ set by \cite{Belczynski02}.

Events from the peak of the probability distribution function (see Fig. \ref{fig_pdf}) dominate short observation
times. As observation time increases, the rarer events at both high and low redshift become more numerous. The high-$z$
events will be buried in detector noise, and only potentially detectable by cross-correlation between co-located
detectors \citep{Maggiore00}. The background component from more luminous events, $z \lesssim 0.4$, could be detected
by Advanced LIGO as individual burst events. When we apply the PEH algorithm to these types of data, the PEH
population, $C$, will consist mostly of the rarer events at the top edge of Fig. \ref{fig_injected}.

\begin{figure}
\includegraphics[width=84mm]
{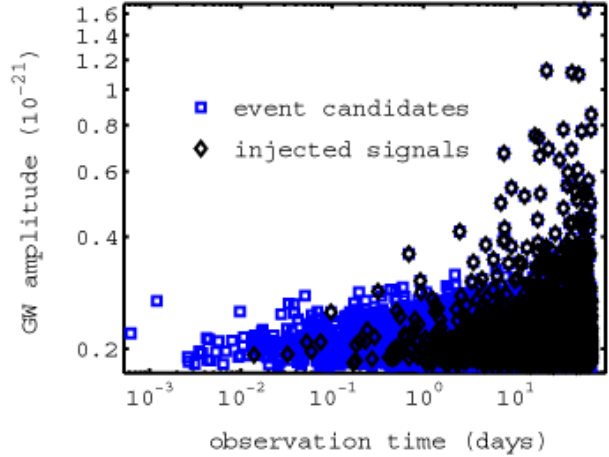}
 \caption{The candidate population of BBH for 4 months of observation time -- around 37,000 events --
 represented by grey squares. Of these candidates, the 2173 events that correspond to injected signals are
shown as black diamonds. The remainder are classed as false alarms.}
 \label{fig_candandinjected}
\end{figure}

Figure \ref{fig_candandinjected} displays the output of our simulation pipeline for an observation time of four months,
constituting 37,387 events. Of this population, 2173 candidates were identified as injected signals -- the remainder
are false alarms. At early observation times noise events dominate and the false alarm rate is high. As observation
time increases, a greater proportion of the candidate signals are GW burst events.

\subsection{The PEH population}

\begin{figure}
\includegraphics[width=84mm]
{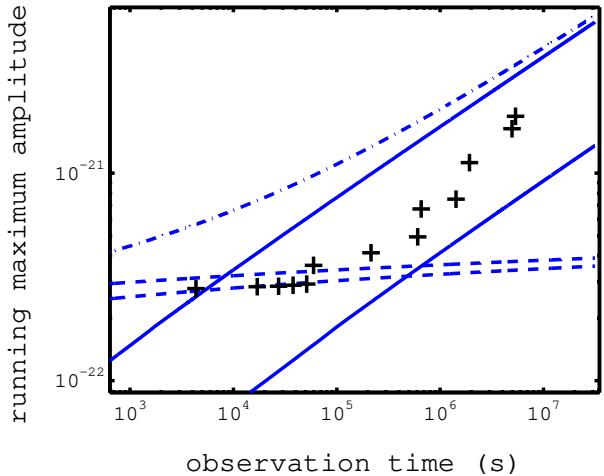} \caption{The PEH population of BBH for 4 months of observation time at advanced LIGO
 sensitivity are shown by crosses, with the solid lines representing the associated 90\% amplitude PEH thresholds
 (see Fig \ref{fig_peh_sources}). In addition, we show the 90\% Gaussian noise amplitude PEH curves (dashed lines),
  which illustrate the temporal amplitude evolution of Gaussian detector noise. The dot-dashed line represents the
  absolute amplitude upper limit at 95\% confidence for signal + noise.}
  \label{fig_ALIGO_PEH}
\end{figure}

Figure \ref{fig_ALIGO_PEH} shows the PEH population of BBH events extracted from one 4-month data segment. Each point
represents the maximum amplitude of events observed as a function of observation time. Also shown is the BBH inspiral
amplitude PEH of Fig. \ref{fig_peh_sources} and the Gaussian noise amplitude PEH threshold, which models the
progressive maximum amplitude growth of the Gaussian detector noise used in our simulation (see section 4.2). The small
gradient of the noise PEH in comparison with the astrophysical PEH curve highlights the fact that the temporal
evolution of the Gaussian detector noise, highly dependent on the sampling frequency, is much slower, a result of the
low probability of events in the tail of the distribution \citep{Coward_NSM05}. Similarly, the narrower 90\% threshold
for the Gaussian-noise PEH is also a result of tighter constraints on the amplitude distribution for the Gaussian noise
model.

For early observation times ($\lesssim$ 30,000 s) the PEH population is dominated by false alarms, lying
close to the Gaussian noise PEH. As observation time increases, the astrophysical PEH population begins to
dominate. An additional threshold, shown by the dot-dashed line, is constructed by combining the null-PEH
curves from both populations. This threshold is the 95\% amplitude upper limit for our BBH inspiral PEH
population -- this curve represents the result of obtaining maximum amplitudes for both the noise and
sources.

\section{The Log N - Log A distribution}

We now want to determine the key astrophysical parameter, the source local rate density $r_{0}$, from data such as that
presented in the previous section. First we will consider the conventional method of determining source rate densities
based on the brightness distribution of sources. This will allow us to quantify the effectiveness of the PEH in
determining the local rate density of BBH events. The log N--log P source count distribution
\citep{Guetta05,Totani99,Schmidt01} is the number of events $N (>\hspace{0.5mm}P)$, of luminosity $L$ and peak flux
$P$, within a maximum redshift, $z_{\mathrm{max}}(L, P_{\mathrm{lim}})$, recorded by a detector of flux limit
$P_{\mathrm{lim}}$. Such a distribution can be fitted to a predicted log N--log P curve to obtain rate estimates or to
constrain the luminosity function.

For our standard-candle GW population, we convert peak flux to a maximum GW amplitude, $A$, yielding a log N--log A
distribution of the form:

\begin{equation}
N(> A)=\int_{0}^{z (L, A)} \tau_{\mathrm{0}} \frac{\mathrm{d}R}{\mathrm{d}z}  \mathrm{d}z  \,
\end{equation}

\noindent where $\tau_{\mathrm{0}}$ represents the observation time and the differential event rate,
$\mathrm{d}R/\mathrm{d}z$, is given by equation \ref{drdz}. In comparison with the log N--log P distribution, which has a
gradient of $-3/2$ under a Euclidean geometry, the log N--log A has a slope of $-3$.  The curve includes a noise
component which approximates the average contribution from detector noise and is scaled by a factor 0.4, the mean value
of the antenna response function for a single GW detector \citep{Finn93}. By fitting to 100 synthetic data sets, we find
these approximations introduce a systematic error of $\pm 6\% $ to the final estimates.

To fit the candidate event population, $E$, against the log N--log A curve, we consider two scenarios: firstly an
idealized case, in which the detector has correctly resolved all the injected events in $E$; secondly a suboptimum case,
in which the data consists of both injected signals and false alarms. To implement the first scenario we eliminate any
false alarms by including only event triggers that correspond to injected signals. For the second scenario we use the
whole candidate population of event triggers and false alarms. In section 8 we will apply the PEH method to both
scenarios, thereby allowing a direct comparison.

\begin{table}
\begin{centering}
\begin{tabular}[scale=1.0]{ccc}
  \hline
Data stream      &SFR model            & Estimate of $r_{0}$                         \\
   &              &                    (in units of $\tilde{r}_{0}$)         \\
  \hline
1        &SF2                         & $1.00 \pm 0.07 $        \\
1        & constant                   & $1.74 \pm 0.11 $         \\
1        &SF1                         & $0.99 \pm 0.07 $        \\
1        &SH                          & $1.46 \pm 0.10 $         \\
  \hline
2        &SF2                         & $1.01 \pm 0.07 $        \\
2        & constant                   & $1.67 \pm 0.10 $        \\
2        &SF1                         & $0.99 \pm 0.07 $        \\
2        &SH                          & $1.48 \pm 0.10 $        \\

\end{tabular}

\caption[waveform parameters]{The results of least squares fitting to the log N -- log A distributions of two
independent data steams, each representing 4 months of observation time. The data consists of 2273 events for data
steam 1 and 2173 events for data stream 2. The results for data stream 1 are shown in Fig \ref{fig_logNlogA1}. The data
streams represent the output of a perfect detector -- all false alarms have been dismissed. The SFR model is the one
used in the fit; the estimated ranges of $r_{0}$ are given at 90\% confidence.}
\end{centering}
\label{table_logNlogA1}
\end{table}

\begin{table}
\begin{centering}
\begin{tabular}[scale=1.0]{ccc}
  \hline
Data stream    &SFR model          & Estimate of $r_{0}$                         \\
  &             &                           (in units of $\tilde{r}_{0}$)   \\
  \hline
1        &SF2                         & $1.36 \pm 0.08$        \\
1        &constant                    & $1.96 \pm 0.13 $         \\
1        &SF1                         & $1.35 \pm 0.08 $        \\
1        &SH                          & $1.80 \pm 0.12 $         \\
  \hline
2        &SF2                         & $1.28 \pm 0.08 $        \\
2        &constant                    & $1.86 \pm 0.13$        \\
2        &SF1                         & $1.27 \pm 0.08 $        \\
2        &SH                          & $1.77 \pm 0.13 $        \\
\end{tabular}

\caption[waveform parameters]{The same as for Table 1, but for data with a high false alarm rate. The data consists of
38241 events for data stream 1 and 37873 events for data stream 2. The results for data stream 1 are shown in Fig
\ref{fig_logNlogA2}.}
\end{centering}
\label{table_logNlogA2}
\end{table}

\begin{figure}
\includegraphics[width=84mm]
{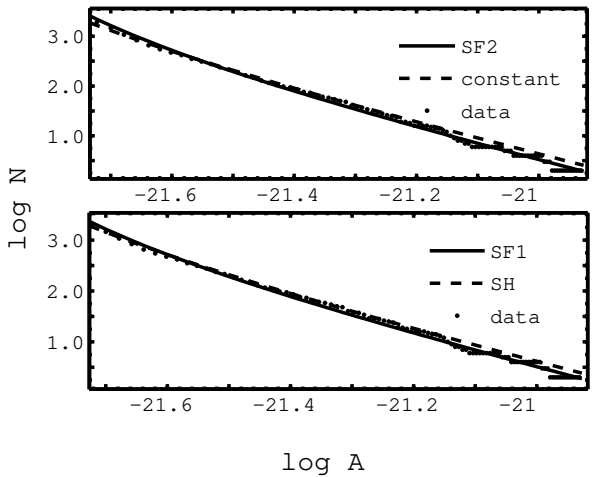} \caption{A least squares fit to simulated linear-linear GW data, using predicted brightness distributions
based on different SFR models, with a single free parameter, $r_{0}$. The data are candidate events from an event
population that correspond to injected signals -- this represents an idealized data set, in which we assume all false
alarms have been dismissed by the detector. The top panel shows the result of fits based on SF2 and a constant
(non-evolving) model; the bottom panel shows fits based on SF1 and SH. The estimates obtained using this fitting
procedure are given in Table 1.}
  \label{fig_logNlogA1}
\end{figure}

Figure \ref{fig_logNlogA1} shows the results of fitting against an idealized data set. For data streams 1 and 2 we
obtain estimates of $( 1.00 \pm 0.07) \tilde{r}_{0}$  and $( 1.01 \pm 0.07) \tilde{r}_{0}$ within 90\% confidence using
a non-linear regression model based on SF2, the model used to simulate our data. Table 1 shows that the estimates
obtained using fits based on four different SFR models recover the true rate to within an average of 46\%. These
estimates represent the upper limits obtainable by applying the log N -- log A method to our simulated data set. We
note that the choice of SFR model introduces a bias in our estimates of $r_{0}$. These biases result from the SFR model
dependence of the all-sky event rate, $R(z)$, as illustrated in Figure \ref{fig_rz_low}.

\begin{figure}
\includegraphics[width=84mm]
{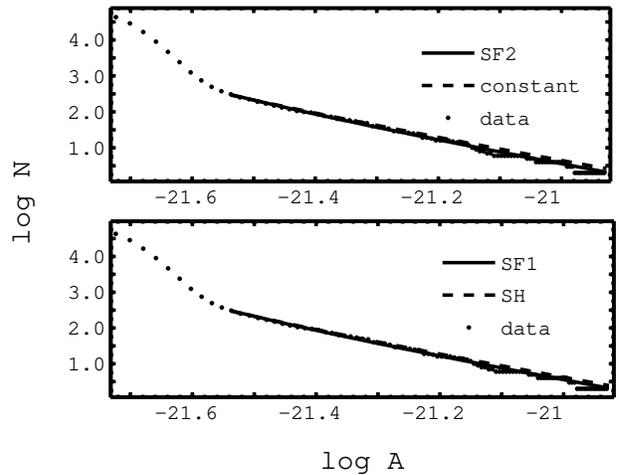}
 \caption{The non-linear regression models of Fig. \ref{fig_logNlogA1}, are fitted to the full
candidate population, consisting of both injected events and false alarms. We fit to the data above a threshold of log
A = -21.55. The estimates obtained using this fitting procedure are given in Table 2.
 }
  \label{fig_logNlogA2}
\end{figure}

Figure \ref{fig_logNlogA2} highlights the effect of false alarms on the data sample. At low amplitude, the distribution
is augmented by the noise transients resulting in overestimates of the local rate density. We therefore use a
thresholding procedure to dismiss all events below log A = -21.55. Using a fit based on the SF2 model, we obtain
estimates of $( 1.36 \pm 0.08) \tilde{r}_{0}$ using data stream 1  and $( 1.28 \pm 0.08) \tilde{r}_{0}$  for data
stream 2. The estimates obtained using fits based on the four different SFR models are shown in Table 2. The
consistency of each pair shows that the statistical errors are small. In addition to the biases resulting from the
different SFR models used, for this suboptimum case there is a significant bias due to the effect of false alarms in
the distribution. In comparison, one might expect the PEH method, which considers only the most energetic events as a
function of time, should be less sensitive to false alarms due to noise, but may have less precision since not all
events are utilized. This is considered in the analysis below.

\section{Fitting to the PEH population}

The dependence of the PEH on $r_{0}$, as demonstrated in Fig. \ref{fig_peh_lrd}, provides a means of fitting the PEH
curves to a candidate PEH population, $C$, and estimating $r_{0}$. However, an obstacle to this procedure is
highlighted in Fig. \ref{fig_ALIGO_PEH}, which shows that  $C$ is dominated by detector noise at early observation
times ($t$ $\lesssim$ 30,000\hspace{0.5mm}s). This reduces the samples available to fit to the amplitude PEH curves.

To reduce the inclusion of false alarms in the fitting procedure, we only include the sample of $C$ above
30,000\hspace{1.0mm}s. This threshold corresponds to the change in gradient of the PEH distribution as injected events
become dominant over the Gaussian noise components at long observation times (see Fig. \ref{fig_ALIGO_PEH}). This
choice of threshold will vary for different astrophysical populations, depending on the emission energies and rates.
After applying this threshold, we expect a total PEH population, $C$, of about 7 -- 11 events for four months of
observation time. We can compensate for any loss of events by this thresholding procedure, by utilizing the signature
of the PEH to increase our sample space as discussed below.

\begin{table}
\begin{centering}
\begin{tabular}[scale=1.0]{cccc}
  \hline
Test     &Subset                 & Sample space of  & KS Probability\\
          &combination           & test samples      & $P(>Z_{\mathrm{n}})$\\
  \hline
1        &1 $\times$ 4 month     & 8, 8       & 0.36 \\
2        &2 $\times$ 2 month   & 10, 11       & 0.59 \\
3        &3 $\times$ 1.3 month     & 17, 18       & 0.74  \\
4        &4 $\times$ 1 month  & 22, 21       & 0.81  \\
\end{tabular}
\end{centering}
\label{table-ks-parameters} \caption[waveform parameters]{The results obtained from 2D Kolmogorov-Smirnov tests between
two 4-month samples of BBH data, extracted from the same simulated data stream. Each sample is split into different
numbers of subsets of equal duration and recombined to increase the sample size. We see that the KS statistic improves
with sample space for these combinations.}
\end{table}

We can increase the sample space of $C$ by splitting our candidate event population, $E$, into $l$ subsets of equal
observation time. By applying the PEH filter to each subset and recombining, we form a more highly populated sample
$\widetilde{C}$ -- the temporal duration of which will now correspond to the length of each individual subset.

The available subset configurations are given by $S_{i},\hspace{0.5mm} i = 1,...,l$. As the PEH is
independent of when the detector is switched on, the PEH signature imprinted within each subset, $S_{i}$,
will be set by the subset's duration. When we combine all $S_{\mathrm{i}}$, to form $\widetilde{C}$, the
individual signatures will be subsequently imprinted within the overall sample space. This useful procedure
enables us to increase the statistical sample.

When splitting and recombining $C$, there is a delicate balance between improving the statistics by increasing the
sample space, and reducing it by decreasing the duration. The most efficient way to increase the population of $C$,
whilst retaining the embedded statistical signature of the PEH can be determined using a two-dimensional
Kolmogorov-Smirnov test \citep{FF87}. This test determines the probability of two sets being drawn from the same
population distribution.

We calculated the two-dimensional Kolmogorov-Smirnov probabilities $P^{\mathrm{KS}}$ obtained using two consecutive
4-month samples taken from the same simulated data stream. A high value of $P^{\mathrm{KS}}$ will provide evidence of
the statistical compatibility of two samples. Table 3 shows that a KS probability of 81\% was obtained by splitting the
data into four 1-month subsets and recombining. We therefore use this configuration to produce $\widetilde{C}$, the PEH
data to which we will apply our fitting procedure.

It should be noted however, that the best choice of $l$, corresponding to the sample configuration with the
strongest statistical PEH signature, may vary for different candidate event populations, $E$, depending on
the observation time. For example, in performing the same test on three months of data we found that a
larger $P^{\mathrm{KS}}$ was obtained using $l = 3$ than for $l = 4$ -- the result of a loss in the PEH
signature for samples of shorter temporal duration. For this paper, to illustrate the PEH technique and
avoid any additional complications, we maximize the sample space by using a 4-month sample of data.

To constrain the local rate density we apply a least-squares fit to the candidate population $\widetilde{C}$ using an
amplitude PEH curve as a linear regression model with free parameter $r_{0}$. When equation \ref{eq_peh} is fitted to the
data it is necessary to determine the value of $\epsilon$ that correctly estimates $r_{0}$. This value, equal to 0.36,
was experimentally verified by fitting to 1000 synthetic data sets using $\epsilon$ as the only free parameter.

\begin{figure}
\includegraphics[width=84mm]
{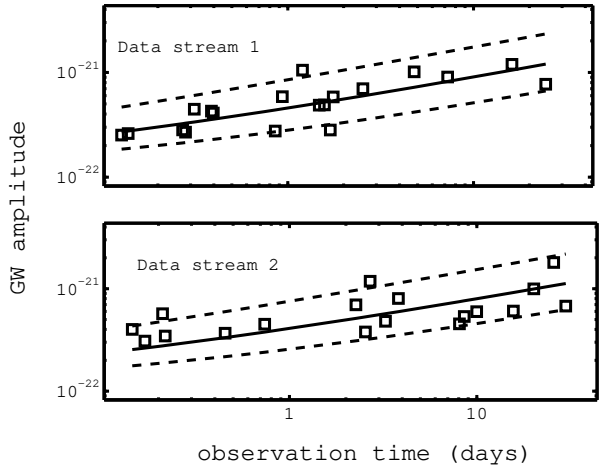} \caption{A least squares fit to simulated GW data using a 36\% PEH curve based on SF2 as a non-linear
regression model. The data correspond to an idealized situation in which all false alarms have been dismissed. Using a
PEH fit based on SF2 we obtain an estimate of $( 1.26 \pm 0.56) \tilde{r}_{0}$ for data stream 1 and $( 0.84 \pm 0.53)
\tilde{r}_{0}$ for the data stream 2. As an additional test of the PEH model, the dashed lines show that the 90\% PEH
thresholds are a good fit to the data.}
  \label{fig_PEH_fit1}
\end{figure}

\begin{figure}
\includegraphics[width=84mm]
{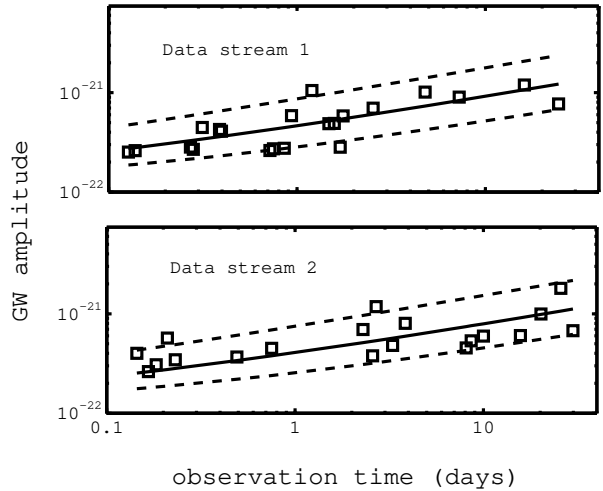} \caption{The same fitting function as used in  Figure \ref{fig_PEH_fit1} is applied to data which include
both injected events and false alarms. We obtain an estimate of $( 1.43 \pm 0.61) \tilde{r}_{0}$ for data stream 1 and
$( 0.83 \pm 0.51) \tilde{r}_{0}$ for the data stream 2.  We again fit the 90\% PEH thresholds to the data, shown by the
dashed lines.}
  \label{fig_PEH_fit2}
\end{figure}

\section{RESULTS}

\begin{table}
\begin{centering}
\begin{tabular}[scale=1.0]{cccc}
  \hline
Data stream      &SFR model      &Events      & Estimate of $r_{0}$                         \\
    &              &            &               (in units of $\tilde{r}_{0}$)  \\
  \hline
1        &SF2                    &19     & $1.26 \pm 0.56 $        \\
1        &constant               &19     & $1.43 \pm 0.63 $        \\
1        &SF1                    &19     & $1.25 \pm 0.56 $        \\
1        &SH                     &19     & $1.37 \pm 0.60$         \\
  \hline

2        &SF2                    &18     & $0.84 \pm 0.53$        \\
2        &constant               &18     & $0.94 \pm 0.57 $        \\
2        &SF1                    &18     & $0.83 \pm 0.52 $        \\
2        &SH                     &18     & $0.92 \pm 0.56 $        \\
\end{tabular}

\caption[waveform parameters]{The results of least squares fitting to the PEH distributions of two independent data
steams, each representing 4 months of observation time. The data represents that of an idealized situation in which all
false alarms have been dismissed. A PEH fit to data stream 1 is shown in Fig \ref{fig_PEH_fit1}. The SFR model is the
one used in the fitting function. The numbers of events used in each fit are shown along with the estimates of $r_{0}$
given at 90\% confidence.}
\end{centering}
\label{table_peh1}
\end{table}

\begin{table}
\begin{centering}
\begin{tabular}[scale=1.0]{cccc}
  \hline
Data stream    &SFR model    &Events      & Estimate of $r_{0}$                         \\
  &             &             &             (in units of $\tilde{r}_{0}$)   \\
  \hline
1        &SF2                   &21        & $1.43 \pm 0.61 $        \\
1        &constant              &21        & $1.61 \pm 0.71 $         \\
1        &SF1                   &21        & $1.42 \pm 0.63 $        \\
1        &SH                    &21        & $1.55 \pm 0.69$         \\
  \hline
2        &SF2                   &19        & $0.83 \pm 0.51 $        \\
2        &constant              &19        & $0.94 \pm 0.55  $        \\
2        &SF1                   &19        & $0.82 \pm 0.51 $        \\
2        &SH                    &19        & $0.91 \pm 0.54  $        \\
\end{tabular}

\caption[waveform parameters]{The same as for Table 1, but for data with a high false alarm rate. A PEH fit to data
stream 1 is shown in Fig \ref{fig_PEH_fit2}.}
\end{centering}
\label{table_peh2}
\end{table}

In this section we present the results obtained by least-squares fitting to candidate PEH populations $\widetilde{C}$
using the 36\% amplitude PEH curve as a linear regression model with free parameter $r_{0}$. We use the same two sets
of data as in section 5 and consider again two different scenarios: firstly, an idealized case in which the detector
operates at high efficiency and has correctly dismissed all false alarms, and secondly, a suboptimum case in which data
include false alarms.

Figure \ref{fig_PEH_fit1} shows the results of a non-linear fit to both data streams for the idealized case in which
all false alarms have been dismissed. This fit uses a PEH curve based on SF2, the model used to generate the data, as a
non-linear regression model and yields estimates of $( 1.26 \pm 0.56) \tilde{r}_{0}$ for data stream 1 and $( 0.84 \pm
0.53) \tilde{r}_{0}$ for data stream 2. As an additional test of the PEH model, this figure shows that the data is well
constrained by the 90\% amplitude PEH thresholds.

Estimates of $r_{0}$ obtained using PEH curves based on all SFR models are shown in Table 4. We see that we obtain mean
estimates to within around a factor of 1.5 for data streams 1 and 2. As a result of the smaller data set used in the
PEH fitting procedure, estimates can be sensitive to the distribution of data. This effect is highlighted in Figure
\ref{fig_PEH_fit1} which shows that the PEH distribution of data stream 1 tends towards the upper PEH threshold,
producing higher estimates of $r_{0}$.

The uncertainties in these estimates are greater than those obtained using the brightness distribution. For example,
using a fit based on SF2, the brightness distribution recovered the true rate within 7\% and 8\% for data streams 1 and
2, while the equivalent estimates obtained using the PEH fit were within 82\% and 69\%. We note however, that the
results obtained using the PEH method are not as prominently effected by bias due to different SFR models as those
obtained using a brightness distribution.

In Figure \ref{fig_PEH_fit2} we see the result of a least-squares fit to data which contains both the injected events
and false alarms.  The numerical estimates determined by applying the PEH fitting procedure to these data are presented
in Table 5 for both data streams 1 and 2. The mean estimates for both data streams are within a factor of 1.5 of the
true value of $r_{0}$.

To determine the effect of false alarms on our estimates it is useful to eliminate the effects of using different SFR
models in our fitting procedures. This can be best achieved by looking at the results obtained using SF2, the model used
to produce the candidate data streams. We see that a PEH fit based on SF2 yields estimates of $( 1.43 \pm 0.61)
\tilde{r}_{0}$ for data stream 1 and $( 0.83 \pm 0.51) \tilde{r}_{0}$ for data stream 2. We see that the inclusion of
false alarms has resulted in a 24\% and a 3\% decrease in accuracy for data streams 1 and 2 respectively. In comparison,
estimates using an SF2 model to fit to the brightness distribution degraded by up to 44\% for data stream 1 and 35\% for
data stream 2.

These results imply that the PEH method is less prone to errors due to the inclusion of false alarms. This outcome arises
because the PEH distribution is composed of the most energetic events as a function of observation time, thereby omitting
most false alarms. We note however that the PEH method has a lower resolution than that of the brightness distribution --
a direct result of the smaller data sets used in the fitting procedure. This obvious disadvantage will be discussed in
the next section.

Tables 6 and 7 show the estimates obtained by combining brightness distribution and PEH data. When compared to the
results obtained using the brightness distribution in Tables 1 and 2, we see that including the PEH data improves the
estimates by at least 1-2\% in most cases. In addition, these results indicate that the PEH method can be employed as an
additional test of consistency.

\begin{table}
\begin{centering}
\begin{tabular}[scale=1.0]{ccc}
  \hline
Data stream      &SFR model            & Estimate of $r_{0}$                         \\
    &              &                     (in units of $\tilde{r}_{0}$        \\
  \hline
1        &SF2                         & $1.00 \pm 0.06 $        \\
1        & constant                   & $1.73 \pm 0.10 $         \\
1        &SF1                         & $0.99 \pm 0.06 $        \\
1        &SH                          & $1.45 \pm 0.09$         \\
  \hline
2        &SF2                         & $1.00 \pm 0.06 $        \\
2        & constant                   & $1.64 \pm 0.09 $        \\
2        &SF1                         & $0.98 \pm 0.06 $        \\
2        &SH                          & $1.46 \pm 0.09 $        \\
\end{tabular}

\caption[waveform parameters]{The constraints on the true local rate density, $r_{0}$, obtained by combining the
results of least squares fitting to the log N -- log A distributions (Table 1) and the PEH distributions (Table 4) of
the two independent data steams. The data sets represent the output of a perfect detector -- all false alarms have been
dismissed. As previously, the SFR model is the one used in the fit; the estimated ranges of $r_{0}$ are given at 90\%
confidence.}
\end{centering}
\label{table_logNlogA_peh1}
\end{table}

\begin{table}
\begin{centering}
\begin{tabular}[scale=1.0]{ccc}
  \hline
Data stream    &SFR model          & Estimate of $r_{0}$                         \\
  &             &                       (in units of $\tilde{r}_{0}$)      \\
  \hline
1        &SF2                         & $1.36 \pm 0.07$        \\
1        &constant                    & $1.94 \pm 0.12 $         \\
1        &SF1                         & $1.35 \pm 0.07 $        \\
1        &SH                          & $1.79 \pm 0.11 $         \\
  \hline
2        &SF2                         & $1.26 \pm 0.07$        \\
2        &constant                    & $1.81 \pm 0.12 $        \\
2        &SF1                         & $1.25 \pm 0.07 $        \\
2        &SH                          & $1.72 \pm 0.12 $        \\
\end{tabular}

\caption[waveform parameters]{The same as for Table 6, but this time we combine the estimates of Tables 2 and 5 for data
with a high false alarm rate.}
\end{centering}
\label{table_logNlogA_peh2}
\end{table}

\section{Conclusions }

We have shown that the PEH filter allows an independent estimate of the rate density of BBH coalescence events detected
by advanced GW detectors. The main results and their limitations are summarized below:

\begin{enumerate}

\item For a candidate population of BBH with Galactic inspiral rate ${\cal R}_{\rm gal}^{BBH}\sim
30\hspace{1.0mm}\mathrm{Myr}^{-1}$, a fit to 4 months of interferometer data was sufficient to obtain estimates of
$r_{0}$ to within a factor of 2 at the 90\% confidence level.

By applying both brightness distribution and PEH methods to data streams with a high false alarm rate we find that the
brightness distribution method is the more accurate if the detector is operating at high efficiency. For the case in
which the data contains a large proportion of false alarms, we find the PEH fit gives less bias. This is because this
method naturally suppresses low-amplitude false-alarms. Of the two methods, the PEH method has lower resolution due to
the fact that fewer events contribute to the data. The overall performance of the PEH filter suggests that it is
accurate enough to be considered as an additional tool to determine event rate densities, particularly if the detector
is operating at low efficiency. A combination of the PEH and brightness distribution methods provides two independent
estimates of $r_{0}$. The combination of both estimates provides a self consistent test
and also increases the overall precision of the rate estimates. \\

\item One disadvantage of the PEH filter, in comparison with fitting to the amplitude distribution, is that it uses only a small sample
of the overall data set. We are investigating techniques to increase the sample size. Initial results suggest that
applying the PEH filter in both temporal directions can increase the overall PEH population and improve the resolution
and accuracy of the estimates. This technique will be particularly
useful in data sets in which a large event occurs after a comparatively short observation time.\\

\item We note that the value for reference local rate density, $\tilde{r}_{0}$, obtained from the population synthesis calculations of \cite{Belczynski02},
is at the upper end of predictions. We plan to investigate the performance of the PEH method for lower values of
$\tilde{r}_{0}$ for which we expect a smaller number of events in the PEH distribution. Initial calculations suggest
that an order of magnitude decrease in $\tilde{r}_{0}$ will result in a PEH population of around 7 -- 11 events for a
4-month data set rather than the 15 -- 22 events expected for $\tilde{r}_{0}$ used in this study. As discussed
previously, techniques in which we can
increase the PEH sample will be of great importance for astrophysical populations with lower rate densities.\\

\item For the noise levels considered in this study, the PEH fits are only weakly affected by the inclusion of false
alarms, but estimates using a brightness distribution are shown to degrade. This implies that the PEH method may be
most effective when applied to data output from signals that are not well modelled, such as transient burst sources,
for which we expect a substantial number of false alarms to be present. Recent developments in the modelling of GW
emissions from core-collapse supernova (SNe) suggest that the detectable background
population of such sources may be numerous enough to apply the PEH technique \citep{Ott06}.\\

\item The PEH filter, by definition includes the temporal distribution of the sources, hence provides a means
of predicting the energies of future events.\\

\item Our results show that at the detector sensitivity assumed here, the value of $r_{0}$ cannot be separated from
evolution effects, so that different SFR curves create bias in the estimated value of $r_{0}$. For larger spans of data
and more sensitive detectors such as EURO \citep{Saty04} it may be possible to fit the PEH data or the brightness
distribution to determine the entire function $R(z)$.

\end{enumerate}

\section*{Acknowledgments}
This research is supported by the Australian Research Council Discovery Grant DP0346344 and is part of the research
program of the Australian Consortium for Interferometric Gravitational Astronomy (ACIGA). David Coward is supported by
an Australian Postdoctoral Fellowship. The authors are grateful to Peter Saulson of LIGO for an initial reading of the
manuscript and to Bangalore Sathyaprakash who carefully reviewed the paper on behalf of the LIGO Science Committee.
Both these reviews produced insightful feedback and have led to some valuable amendments. The authors also thank the
anonymous referee for some useful suggestions for improving the clarity of the paper.

\label{lastpage}

\end{document}